# Single-pixel coherent diffraction imaging

Meng Li[1], Liheng Bian[1,*], Guoan Zheng[2], Andrew Maiden[3], Yang Liu[4], Yiming Li[1], Qionghai Dai[4], Jun Zhang[1]

[1]School of Information and Electronics & Advanced Research Institute of Multidisciplinary Science, Beijing Institute of Technology, Beijing 100081, China
[2]Department of Biomedical Engineering & Department of Electrical and Computer Engineering, University of Connecticut, Storrs, Connecticut 06269, USA
[3]Department of Electronic and Electrical Engineering, University of Sheffield, Sheffield S1 3JD, UK
[4]Department of Automation, Tsinghua University, Beijing 100086, China
*E-mail: bian@bit.edu.cn

**Complex-field imaging is indispensable for numerous applications at wavelengths from X-ray to THz, with amplitude describing transmittance (or reflectivity) and phase revealing intrinsic structure of the target object. Coherent diffraction imaging (CDI) employs iterative phase retrieval algorithms to process diffraction measurements and is the predominant non-interferometric method to image complex fields. However, the working spectrum of CDI is quite narrow, because the diffraction measurements on which it relies require dense array detection with ultra-high dynamic range. Here we report a single-pixel CDI technique that works for a wide waveband. A single-pixel detector instead of an array sensor is employed in the far field for detection. It repeatedly records the DC-only component of the diffracted wavefront scattered from an object as it is illuminated by a sequence of binary modulation patterns. This decreases the measurements' dynamic range by several orders of magnitude. We employ an efficient single-pixel phase-retrieval algorithm to jointly recover the object's 2D amplitude and phase maps from the 1D intensity-only measurements. No *a priori* object information is needed in the recovery process. We validate the technique's quantitative phase imaging nature using both calibrated phase objects and biological samples, and demonstrate its wide working spectrum with both 488-nm visible light and 980-nm near-infrared light. Our approach paves the way for complex-field imaging in a wider waveband where 2D detector arrays are not available, with broad applications in life and material sciences.**

Complex-field imaging provides both amplitude and phase images of an object, where the amplitude describes the object's transmittance or reflectivity, and the phase characterizes how much the light is delayed through propagation and reveals the object's intrinsic structure [1]-[3]. In many practical applications where object absorption is strong or varies little across the field of view, phase is more favorable with enhanced contrast [2]. While existing optoelectronic detectors (such as charge-coupled devices) acquire light intensity by converting photons to electrons and measure the resulting charge, they cannot follow the electromagnetic oscillation rate of ~$10^{15}$ Hz (or higher at short wavelengths) to record phase [3]. This constitutes the well-known, ill-posed phase problem, which is a fundamental limitation inherent to all complex-field imaging modalities for optical [4]-[6], crystallographic [7], and astronomical applications [8]. Although digital holography [9]-[14] enables phase recovery, the interferometry places high demands on instrument stability, and the spatial and temporal coherence of the associated light sources. This is a particular challenge for imaging at wavelengths in such as X-ray and THz wavebands.

Originating from X-ray crystallography, coherent diffraction imaging (CDI) enables non-interferometric complex-field imaging [15][16]. In a typical CDI setup, a coherent probe beam illuminates an object, which scatters the beam to form a diffraction pattern in the far field, whose intensity is recorded by a photodetector array. Because no lenses are placed between the object and the detector, the recorded data are free from optical aberrations, and the numerical aperture is limited only by the collection angle of the detector array [17]. The complex-field information is reconstructed from the acquired diffraction pattern using phase retrieval algorithms [3]. Benefiting from its non-interferometric nature and the availability of high-power computing resources, CDI has rapidly prospered in the past two decades using either visible light or hard X-rays [18]-[27].

Although the stability and illumination requirements of CDI are less stringent than digital holography, CDI imposes high requirements on array-sensor detection. First, it requires an ultra-high dynamic range to record the diffraction pattern. This is because far-field propagation approximates the Fourier transform following the Fraunhofer diffraction theory [28], and Fourier-domain signals rapidly decay radially from the center to the edge. In some cases the recorded data can span more than 7 orders of magnitude [29][30]. Second, the Fourier-domain signals must fulfill the Shannon oversampling criterion to ensure data redundancy for successful phase retrieval [7][16][30]. In practice, this means that the detector pixel size is inversely proportional to the object size, and that the number of detector pixels must be large enough to capture large scattering angles and thus obtain the required image resolution. Such large-scale and dense detector arrays with high dynamic range are often not available beyond the visible waveband, however, which limits CDI's working spectrum [31][32].

In this work, we report a single-pixel CDI technique that uses a single-pixel detector to realize wide-waveband complex-field imaging. In its workflow, the object being imaged is coherently illuminated by a series of binarized patterns, and a single-pixel detector is placed in the far field to acquire the DC-only components of the diffraction patterns that result. This single-pixel detection scheme lowers the measurements' dynamic range by orders of magnitude, and works for a wide waveband from THz to X-ray. An efficient single-pixel phase-retrieval algorithm is employed to jointly recover the object's 2D amplitude and phase maps from the 1D intensity-only measurements. Fundamentally, the reported technique transforms the inherent challenge of wide-waveband complex-field imaging from one that is coupled to the physical limitations of detector array to one that is solvable through single-pixel phase retrieval.

The single-pixel CDI technique shares its roots with the emerging single-pixel imaging (SPI) paradigm [31][32], which couples high-dimensional information into one-dimensional intensity-only measurements. It differs in the fact that it employs a coherent imaging modality that obtains both the object's amplitude and phase maps. Previous approaches to phase imaging using single-pixel measurements are typically based on phase-shifting interferometry [33]-[37], and suffer from the same limitations of digital holograph that require strong instrument stability and light coherence. In addition, these approaches implement direct matrix inversion to recover the complex-field information, which directly conveys measurement noise onto the recovered results.

In comparison, our gradient-based optimization maintains robustness even for noisy data (see Supplementary Materials for more details).

Besides maintaining robustness to measurement noise, the single-pixel phase retrieval also bypasses the requirement for *a priori* object support constraints in the conventional CDI reconstruction, which involves further experimental costs for accurate calibration of each sample. This feature is enabled by the plurality of illumination modulations, which ensure sufficient data redundancy to remove the phase-retrieval ambiguities inherent to single-measurement forms of CDI [7][38][39]. Although this operation increases acquisition time, it is performed by non-mechanical, ultra-fast modulation of the binarized patterns, and implementation at different wavelengths is further simplified because the binarized modulation contains only two states [40]-[41].

In the following, we will briefly outline the working principle and experimental set-up of single-pixel CDI. Then, we use a prototype system to validate its quantitative phase imaging nature on both calibrated phase objects and biological samples, and demonstrate its wide working spectrum with both 488-nm visible light and 980-nm near-infrared light. Our approach paves the way for complex-field imaging in a wider waveband where 2D detector arrays are not available, with broad applications in life and material sciences.

**Principle of single-pixel CDI**

The single-pixel CDI scheme is presented in Fig. 1. The wide-field illumination is modulated following a series of binarized patterns. A single-pixel detector is placed in the far field to acquire the central DC-only components of the diffraction patterns that result. Following the Fraunhofer approximation [28], the far-field propagation of a light wave approximates its Fourier transform. Therefore, the single-pixel measurement is essentially the DC component of the modulated light field's Fourier spectrum. With the captured 1D intensity signals, we use the developed single-pixel phase retrieval algorithm to recover both the object's amplitude and phase maps.

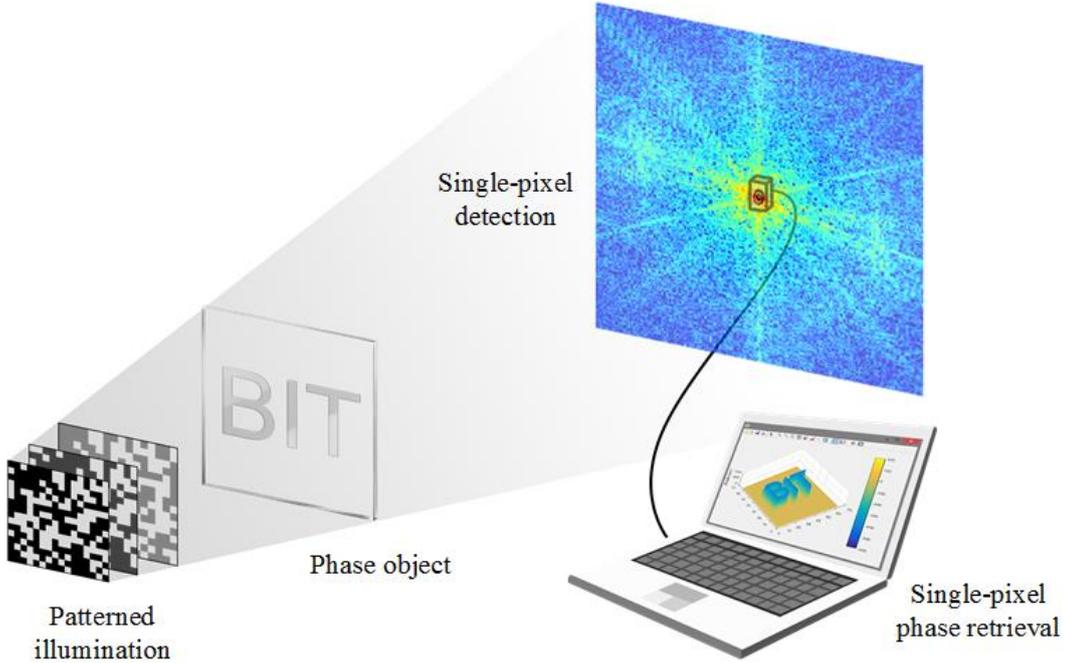

Fig. 1: The scheme of single-pixel coherent diffraction imaging. The wide-field illumination is modulated following a series of binarized patterns. A single-pixel detector is placed in the far field to capture the DC-only components of the diffraction patterns that result. The captured 1D signals are used to recover both the object's 2D amplitude and phase maps using our single-pixel phase retrieval algorithm.

Figure 2(a) presents the forward model describing the measurement formation. Mathematically, the single-pixel measurement can be modeled as

$$I_k = \left| \sum_{(i,j)} \{\mathcal{F}[P_k(i,j) \odot O(i,j)] \odot \delta(i,j)\} \right|^2, \quad (1)$$

where $P_k \in \mathbb{R}^{\sqrt{n} \times \sqrt{n}}$ denotes the $k^{th}$ illumination pattern ($n$ is the total pixel number), $O \in \mathbb{C}^{\sqrt{n} \times \sqrt{n}}$ represents the complex object, $I_k \in \mathbb{R}^{1 \times 1}$ is the $k^{th}$ intensity measurement, $\odot$ stands for entry-wise multiplication, $\mathcal{F}$ represents the two-dimensional Fourier transform, $\delta$ is the impulse function with only the central entry being 1 and other entries being 0, and $\sum_{(i,j)}$ denotes the inner summation of a matrix with subindex $(i,j)$. Essentially, $I_k$ is the intensity of the Fourier spectrum's DC component, and contains coupled amplitude and phase information of the object.

Figure 2(b) shows the single-pixel phase retrieval reconstruction framework. It starts with the initialization of the complex-field image. In our implementation, we use two random matrices following a Gaussian distribution as the initialized amplitude and phase. Then the reconstruction alternates between the spatial and Fourier domains [42]. In each iteration, we multiply the object image with one illumination pattern, and implement the Fourier transform to model propagation to the far-field. The Fourier spectrum is then updated by replacing the zero-frequency amplitude with the square root of the intensity measurement, whilst its phase is kept unchanged. The result is inverse Fourier transformed to obtain an estimate of the modulated target image. The above process repeats sequentially for each illumination pattern and corresponding measurement, until a self-consistent solution of the object's complex-field image is achieved. We note that this alternating projection optimization has lower computational complexity ($O(mnlogn)$) than other phase retrieval methods such as the Wirtinger derivation based ($O(mn^2)$) and the semi-definite programming based ($O(n^3)$) algorithms ($m$ is the number of measurements and $n$ is the number of to-be-reconstructed signals) [3][38][42].

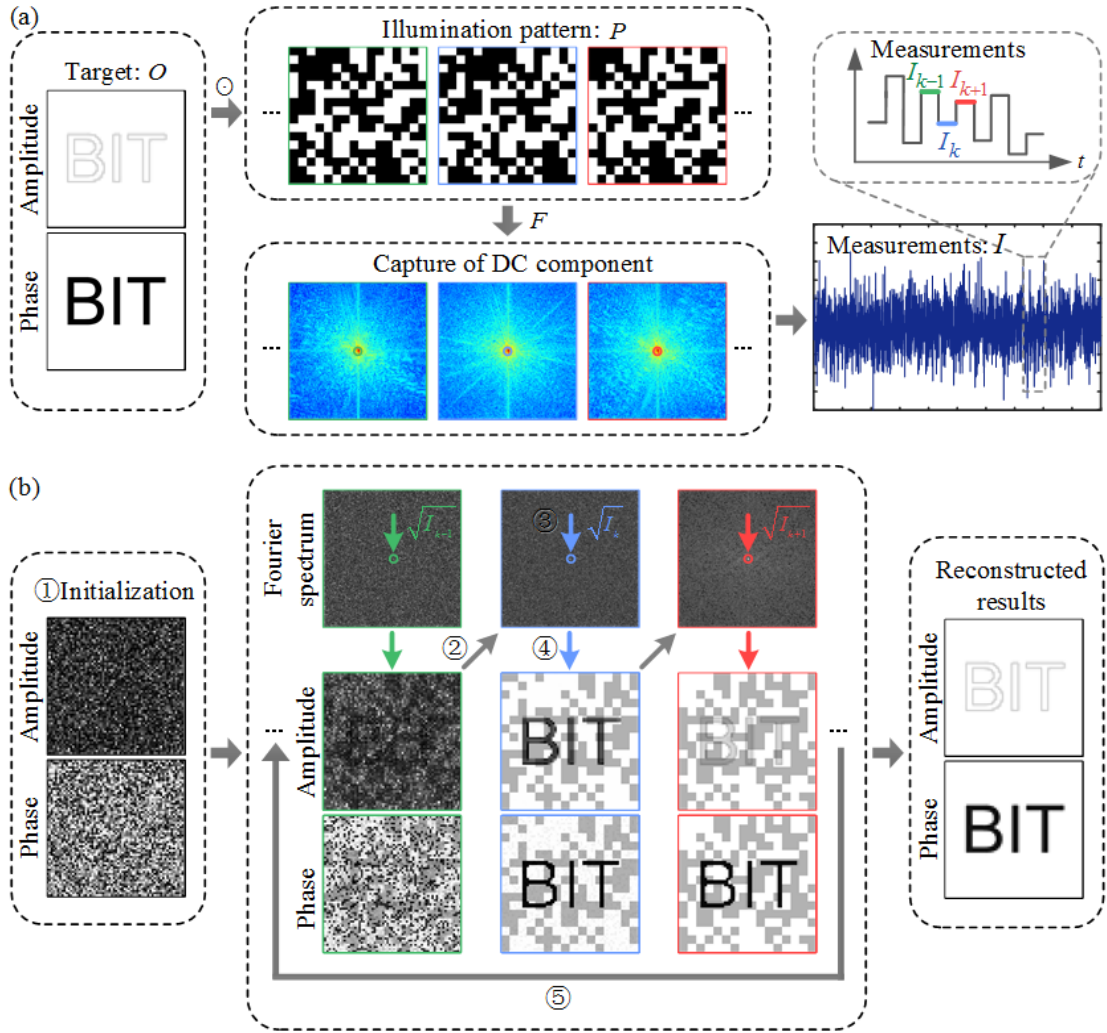

Fig. 2: The forward model and reconstruction framework of single-pixel CDI. (a) The object $O$ is multiplied with the $k^{th}$ illumination pattern $P_k$. The modulated light field $P_k \odot O$ is transformed to the Fourier domain, and only the DC component is recorded as the $k^{th}$ single-pixel measurement $I_k$. (b) The object's complex-field image is computationally recovered by the single-pixel phase retrieval algorithm. Step 1: Initialize the object's complex-field image with random matrices. Step 2: Multiply the object image with one illumination pattern, and implement the Fourier transform to model the far-field propagation. Step 3: Update the Fourier spectrum by replacing the zero-frequency amplitude with the square root of the intensity measurement. Step 4: Apply the inverse Fourier transform to obtain an updated estimate of the modulated target image. Step 5: Repeat steps 2–4 for all the illumination patterns and single-pixel measurements until convergence.

# Experiment results

## Quantitative phase imaging over a wide waveband

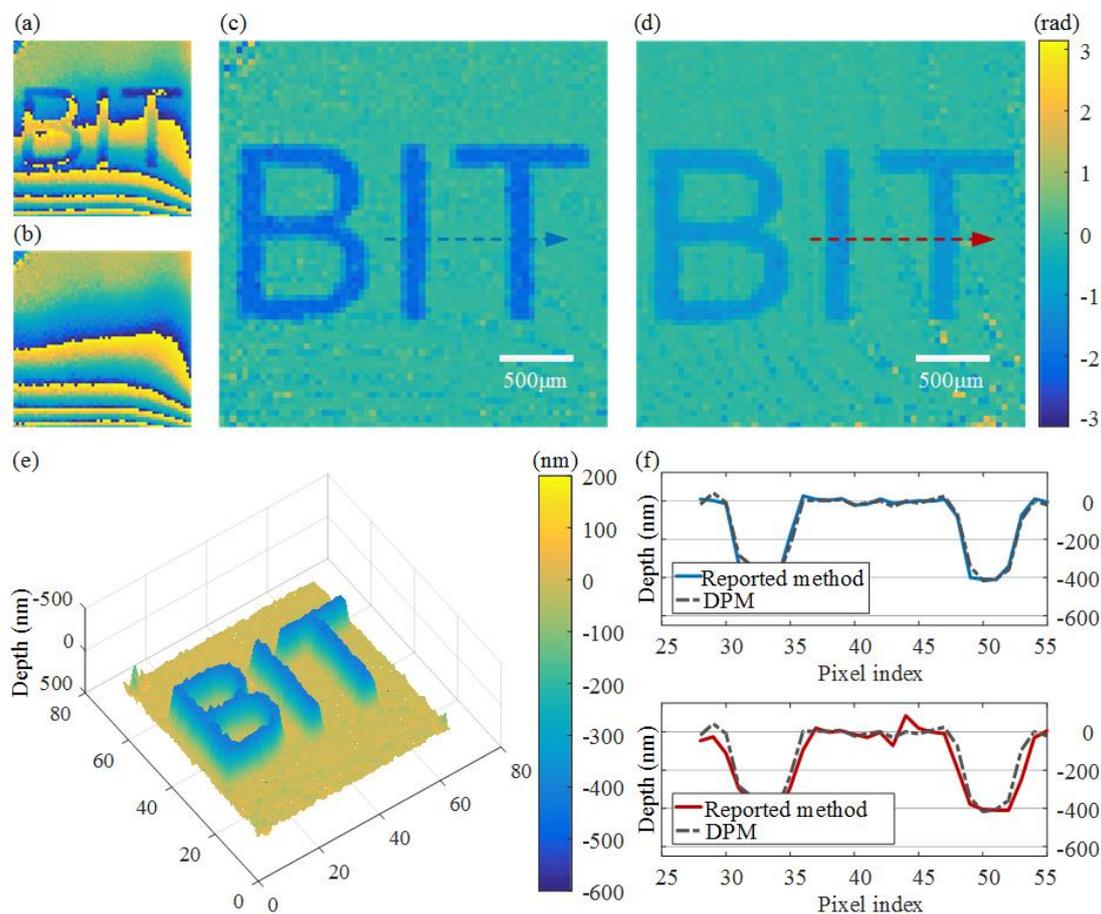

Fig. 3: Quantitative phase imaging results of a transparent phase object using visible light (488nm) and near-infrared light (980nm). (a) Reconstructed phase of the characters "BIT" (uncorrected). (b) Reconstructed background phase. (c) Corrected phase distribution under 488nm light. (d) Corrected phase distribution under 980nm light. (e) Rendered 3D depth distribution of "BIT". (f) Depth profiles along the highlighted lines compared to Diffraction Phase Microscopy (DPM) measurement. Scale bar: 500 μm.

Our first experiment validates the quantitative phase imaging capability of single-pixel CDI over a wide waveband. The object employed in this experiment was a transparent phase target of the characters "BIT", which were etched on a glass slide using reactive ion etching (RIE) and $CHF_3$ gas. The etching depth was 400 nm. We used a 488-nm laser (Coherent Sapphire 488SF 100 mW) and a 980-nm near-infrared laser (CNI MDL-III-980L 200 mW), respectively, as the light source. A digital micromirror device (DMD, ViALUX GmbH V-7001, 400-2500 nm) was employed for binarized modulation. The single-pixel detector (APD100A2, Thorlabs) was placed ~5 m from the object. A 5-μm pinhole was placed in front of the single pixel detector to acquire the DC component of the diffraction pattern. We note that a minimum sampling ratio (the ratio between pattern number and pixel number) of 4 is required for successful single-pixel phase retrieval (see Supplementary Materials for more details). Therefore, to obtain a $64 \times 64$ pixel image, we used 16384 binarized patterns for illumination modulation. The binarized patterns displayed on the DMD switched at a rate of 22 kHz, and the total data acquisition time was ~0.7 seconds.

Figure 3(a) presents the reconstructed phase of the characters "BIT" under 488-nm light, without any correction. We note that there exists a background phase that degrades reconstruction quality. To eliminate the negative influence, we calibrated the background phase by replacing the object with a plane glass slide as shown in Fig. 3(b), and subtracted it from the uncorrected object phase to produce the corrected phase as shown in Fig. 3(c). This subtraction step eliminates the phase aberration caused by detector misalignment from the DC component (refer to the Supplementary Materials for more details). Following the same processing operation, we obtained the corrected phase under 980-nm light as shown in Fig. 3(d).

The relationship between etch depth and phase is given by $h = \frac{\lambda \varphi}{2\pi n}$, where $\lambda$ is the illumination wavelength, $\varphi$ represents the phase, and $n$ denotes the refractive index difference between glass and air ($n = 0.463$ in the experiment) [35]. Using this relationship, the reconstructed 3D depth distribution of the characters "BIT" is shown in Fig. 3(e). The depth profiles along the highlighted line traces are presented in Fig. 3(f). We can see that the recovered phase profiles under both 488-nm light and 980-nm light coincide with the ground-truth depth, as measured by diffraction phase microscopy (plotted as dotted lines).

**Complex-field imaging of biological samples**

We also tested the single-pixel CDI technique with two biological samples, a blood smear shown in Figs. 4(a)-(c) and a pumpkin stem shown in Figs. 4(d)-(f). We used the 488-nm laser for illumination. The image reconstruction size was reset to $128 \times 128$ pixels, requiring 65536 modulated illumination patterns, to reveal more object detail. Because for these smaller objects a shorter imaging distance is enough to meet the far-field propagation condition [2], the detector was placed ~1m away from the samples. For the blood smear sample, the modulated illumination pattern covered an area of $\sim 50\mu m \times 50\mu m$. The reconstructed amplitude map is shown in Fig. 4(a), and the corrected phase is shown in Fig. 4(b) (the Goldstein's branch cut method [43] was used for phase unwrapping). To produce a high-contrast phase map of the sample, we further processed the reconstructed phase following ref. [44] to produce the digital differential interference contrast (DIC) image, as shown in Fig. 4(c). The blood cells are clearly presented. For the pumpkin stem tissue, the illumination pattern covered an area of $100\mu m \times 100\mu m$, and was projected onto the small vascular bundles. The reconstructed amplitude and phase maps are presented in Figs. 4(d) and 4(e), and the reconstructed digital DIC image is presented in Fig. 4(f). The phloem structures are easily recognized.

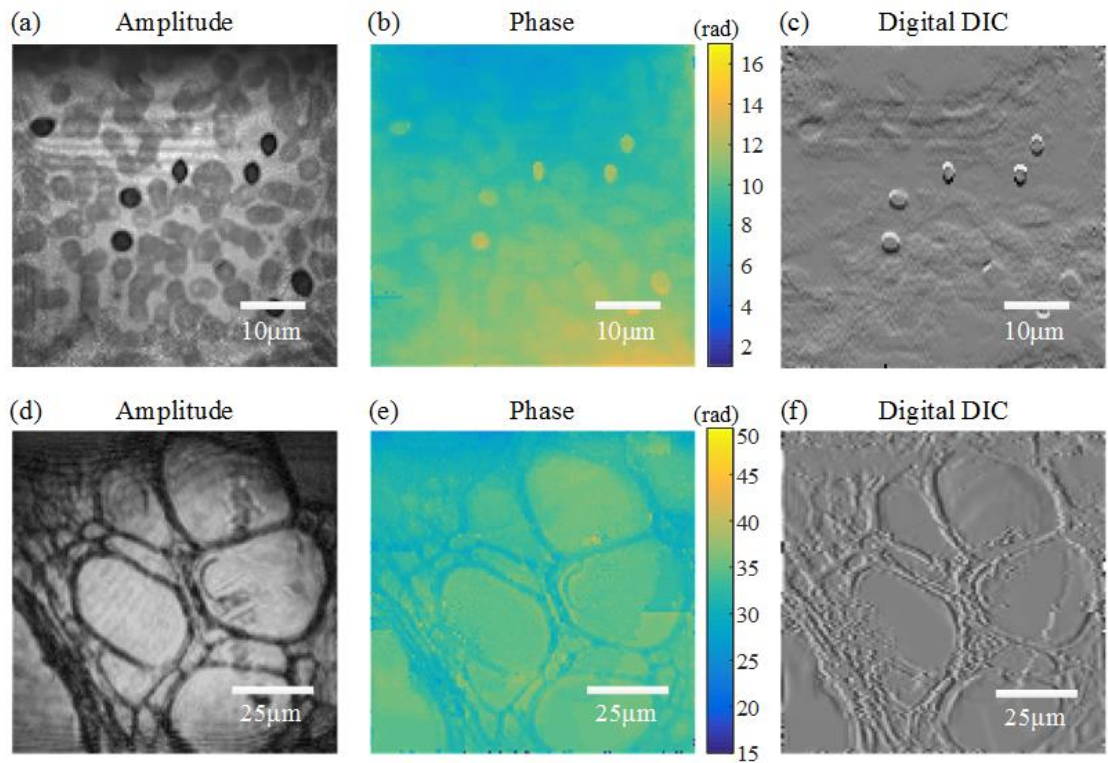

Fig. 4: Complex-field imaging results of biological samples. (a) Reconstructed amplitude map of a blood smear sample. (b) Quantitative phase map of the blood smear. (c) Digital DIC image of the blood smear. Scale bar: 10 μm. (d) Reconstructed amplitude map of a pumpkin stem sample. (e) Quantitative phase map of the pumpkin stem. (f) Digital DIC image of the pumpkin stem. Scale bar: 25 μm.

**Flexible imaging distance**

The third experiment validates the flexibility of our method to different imaging distances. Because the reported technique only acquires the DC component instead of the entire diffraction pattern in the far field, the measurements remain similar for different imaging distances. As demonstrated by the simulation results in the Supplementary Materials, complex-field 2D maps can still be successfully reconstructed even when the diffraction data does not strictly meet the far-field Fraunhofer condition. We experimentally validated this by imaging a monocotyledon stem sample at different imaging distances. The modulated illumination in this experiment covered an area of 100μm × 100μm. The pixel resolution was 128 × 128. The reconstructed amplitude and phase maps at different imaging distances are shown in Fig. 5. We can see that, although there are some aberrations when the imaging distance is short (i.e. 0.01m), the reconstruction quality improves rapidly as the imaging distance increases. Further, as the imaging distance ranges from 0.1m to 1m, the reconstruction quality remains high. We note that the abrupt change in the phase maps of the first two columns is caused by inaccurate phase unwrapping.

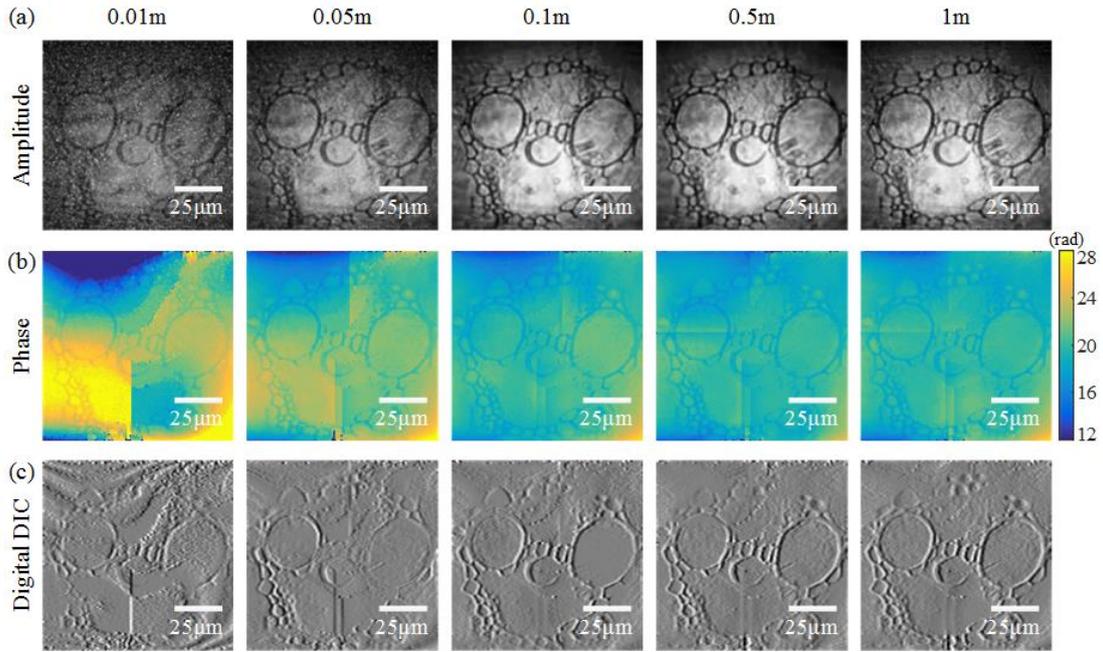

Fig. 5: Complex-field imaging results of a monocotyledon stem sample at different imaging distances (0.01m, 0.05m, 0.1m, 0.5m and 1m). (a) Reconstructed amplitude maps. (b) Quantitative phase maps. (c) Digital DIC images. Scale bar: 25 μm.

**Discussion**

We have demonstrated a single-pixel CDI technique that uses a single-pixel detector to realize complex-field imaging in a wide waveband. Fundamentally, it transforms the inherent challenge of wide-band complex-field imaging from one that is coupled to the physical limitations of optical sensors to one that is solvable through computation. Although the technique implements multiple modulations and acquisitions, it requires no mechanical translation, no high-dynamic-range array detection and no *a priori* object support information. The working spectrum is widely extended owing to the binarized modulation and single-pixel detection. This brings many benefits for imaging applications where large-scale detector arrays are not available, such as THz tomography for security checks [45] and quality inspection [46], where phase images provide intrinsic structure with high contrast [47]. In addition, the DC-only acquisition of the Fourier spectrum ensures high signal-to-noise ratio, allowing weak-light imaging for many applications [48].

Conventionally, the image size of an imaging system is defined by the pixel number of its detector array. The proposed method ultimately reduces the detector array to a single optoelectronic unit; the image size is defined by the patterns projected onto the object by the modulator array. As long as the modulation can be densely implemented at a large scale, complex-field imaging of a wide field can be accomplished. It is this underlying connection that allows our single-pixel CDI prototype to render wide-field images using a single-pixel detector without mechanical scanning.

Drawing connections and distinctions between single-pixel CDI technique and ptychography [24]-[26] helps further clarify the technique's principle of operation. Ptychography can be regarded as a scanning version of CDI, in which the full field is traversed under a focused beam by mechanical translation, and the corresponding far-field diffraction patterns are repeatedly recorded. These images are stitched together to produce a full-field complex-valued image using the phase retrieval reconstruction. It is clear that both single-pixel CDI and ptychography iteratively seek a full-field complex-valued solution that is consistent with multiple intensity-only measurements. With ptychography, the data redundancy for successful phase retrieval is provided by mechanical traversing, and a large portion of illumination overlap is required between adjacent scan. With single-pixel CDI, however, the data redundancy is provided by multiple binarized modulations with no mechanical scanning.

Our single-pixel CDI prototype has not been optimized for imaging speed. At present, the imaging speed is limited by the required number of modulations. This issue can be addressed by using multiple single-pixel detectors or a SPAD array to record multiple spatial frequency signals of the diffraction pattern. As validated by the simulation in the Supplementary Materials, the required modulation number can be effectively reduced by a factor equal to the detection unit number. Another feasible solution is to introduce statistical image priors such as sparsity [23] into the reconstruction, which may allow solution of the phase retrieval problem in a compressive sensing framework [31] that further reduces the required number of modulations and measurements.

Our method can be extended to more general modalities. First, it can be implemented in a passive modality, by moving the spatial light modulator from the illumination path to the detection path. This relieves the thin sample requirement of conventional transmissive-mode imaging systems, because the recovered image in the passive mode depends upon how the complex wavefront exits the object, instead of how it enters [49]. In addition, the recovered complex wavefront can be digitally propagated to other planes along the optical axis, making possible 3D holographic refocusing. Second, the method can be implemented in a reflective modality for high-precision surface inspection, opening the door to multiple industrial applications [50]. Third, a white-light illumination strategy [51] can be introduced into the system to enhance spatial sensitivity, and reduce speckle effects that distort structural details.

**Methods**

In the reported single-pixel CDI technique, the wide-filed illumination is modulated following a series of binarized patterns, and a single-pixel detector is placed in the far field to repeatedly record the DC-only components of the diffraction patterns that result. The object's 2D amplitude and phase maps are computationally reconstructed from the 1D intensity-only measurements.

Recalling the measurement formation model being $I_k = \left|\sum_{(i,j)}\{\mathcal{F}[P_k(i,j)\odot O(i,j)]\odot\delta(i,j)\}\right|^2$, reconstructing the complex-field object image $O(i,j)$ from the single-pixel intensity-only measurements $I$ is essentially a phase retrieval task. Here we present a gradient-based optimization algorithm to tackle the single-pixel phase-retrieval problem. The iterative optimization framework is shown in Fig. 2(b). First, the reconstruction begins with an initial random guess of $O(i,j)$. Second, $O(i,j)$ is multiplied with the $k^{th}$ illumination pattern $P_k(i,j)$ to generate the modulated object $\Psi_k(i,j) = P_k(i,j)\odot O(i,j)$, which is updated in the Fourier domain

as:

$$\mathcal{F}[\Psi_k^{updated}(i,j)] = \mathcal{F}[\Psi_k(i,j)] \cdot [1 - \delta(i,j)] + \sqrt{I_k} \frac{\phi_k}{|\phi_k|} \cdot \delta(i,j), \qquad (3)$$

where $\phi_k = \sum_{(i,j)}\{\mathcal{F}[\Psi_k(i,j)]\odot\delta(i,j)\}$. Third, the target image $O(i,j)$ is updated in the spatial domain as:

$$O^{updated}(i,j) = O(i,j) + \alpha \frac{P_k^*(i,j)}{|P_k(i,j)|_{max}^2} [\Psi_k^{updated}(i,j) - \Psi_k(i,j)], \qquad (4)$$

where $\alpha$ is a tuning parameter that sets the step size of the update. Fourth, the above process repeats sequentially for all the illumination patterns and corresponding measurements until convergence.

The full single-pixel phase-retrieval algorithm is summarized in Alg. 1. We note that in Eq. (2), both $O(i,j)$ and its conjugate $O^*(i,j)$ produce the same single-pixel measurement corresponding to one modulation pattern. Therefore, the reconstruction result is either of these two solutions.

---

**Algorithm 1:** The single-pixel phase-retrieval optimization algorithm.

---

**Input** : Illumination pattern $P \in \mathbb{R}^{\sqrt{n}\times\sqrt{n}\times m}$, measurement sequence $I \in \mathbb{R}^{1\times 1\times m}$, initialization $O \in \mathbb{C}^{\sqrt{n}\times\sqrt{n}}$;

**Output** : Reconstructed target map $O^{updated} \in \mathbb{C}^{\sqrt{n}\times\sqrt{n}}$;

1  $k = 0$;

2  **while** not converge **do**

3  $\quad$ Fourier-domain updating: update $\Psi_k$ following Eq. (3);

4  $\quad$ Spatial-domain updating: update $O$ following Eq. (4);

5  $\quad$ $k \coloneqq k + 1$;

6  **end**

**Acknowledgements**

This work was supported by the National Natural Science Foundation of China (61971045, 61827901) and Fundamental Research Funds for the Central Universities (3052019024). The authors thank Prof. Changhuei Yang for his helpful review.

**Author contributions**

L.B. conceived and supervised the project. L.B. and M.L. performed algorithm derivation. M.L., Y.L. and Y.L. built the setup and conducted the experiments. L.B., M.L., G.Z. and A.M. wrote the paper and discussed with Q.D. and J.Z.

**Competing interests**

The authors declare no competing financial interests.

# Supplementary Materials

## S1: Sampling ratio requirement

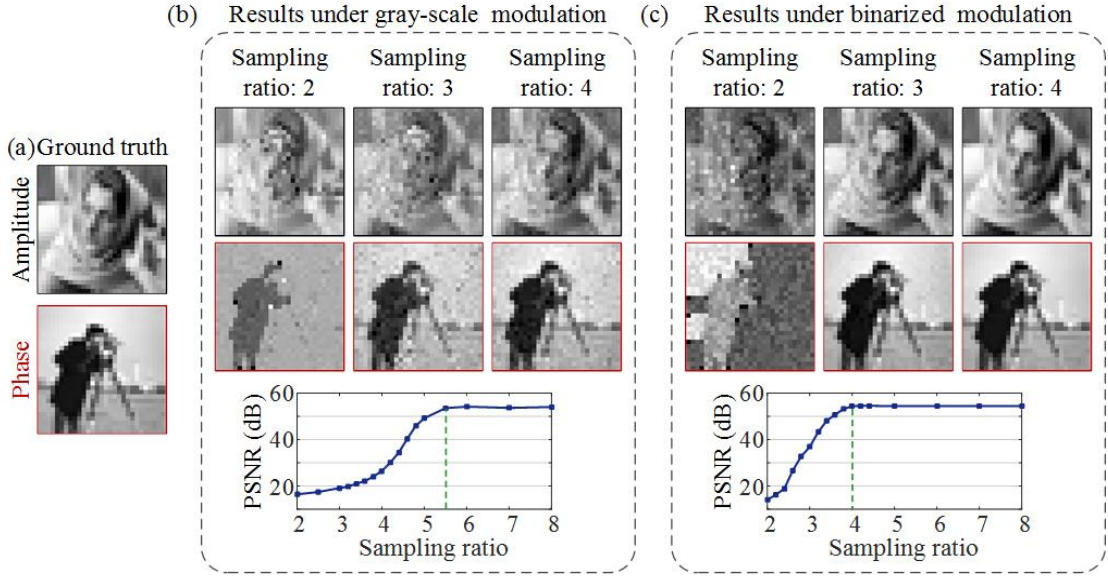

Fig. S1: The reconstructed results under different sampling ratios. (a) The ground-truth amplitude and phase maps of a simulated object, with the pixel number being $32 \times 32$. (b) The reconstructed results under gray-scale modulated illumination. (c) The reconstructed results under binarized modulated illumination.

As described in the Methods section, the wide-field illumination is modulated following a series of binarized random patterns. Here we investigate the sampling ratio requirement for the single-pixel phase retrieval. Sampling ratio is defined as the ratio between the number of illumination patterns and the number of pixels in each pattern.

As shown in Fig. S1(a), we used the 'Barbara' image and 'cameraman' image from the USC-SIPI image database as the ground-truth amplitude and phase maps. Illumination patterns with gray-scale random values (ranging from 0 to 1) and binarized random values (0 and 1) were tested. The measurement sequence of the single-pixel detector was simulated following the measurement formation model Eq. (1). For each sampling ratio, we conducted 50 simulations, and averaged the PSNR of the 50 reconstructed amplitude images to statistically and quantitatively evaluate the reconstruction quality.

Figure S1(b) and S1(c) show the reconstructed results under different sampling ratios using gray-scale and binary patterns, respectively. We can see that as the sampling ratio increases, the image quality improves and approaches an optimum with a sampling ratio of ~5.5 for the gray-scale case and ~4 for the binary case. Considering its low sampling ratio requirement and high speed, we employ binarized modulation in other simulations and experiments, and set the sampling ratio to 4.

## S2: Robustness to measurement noise

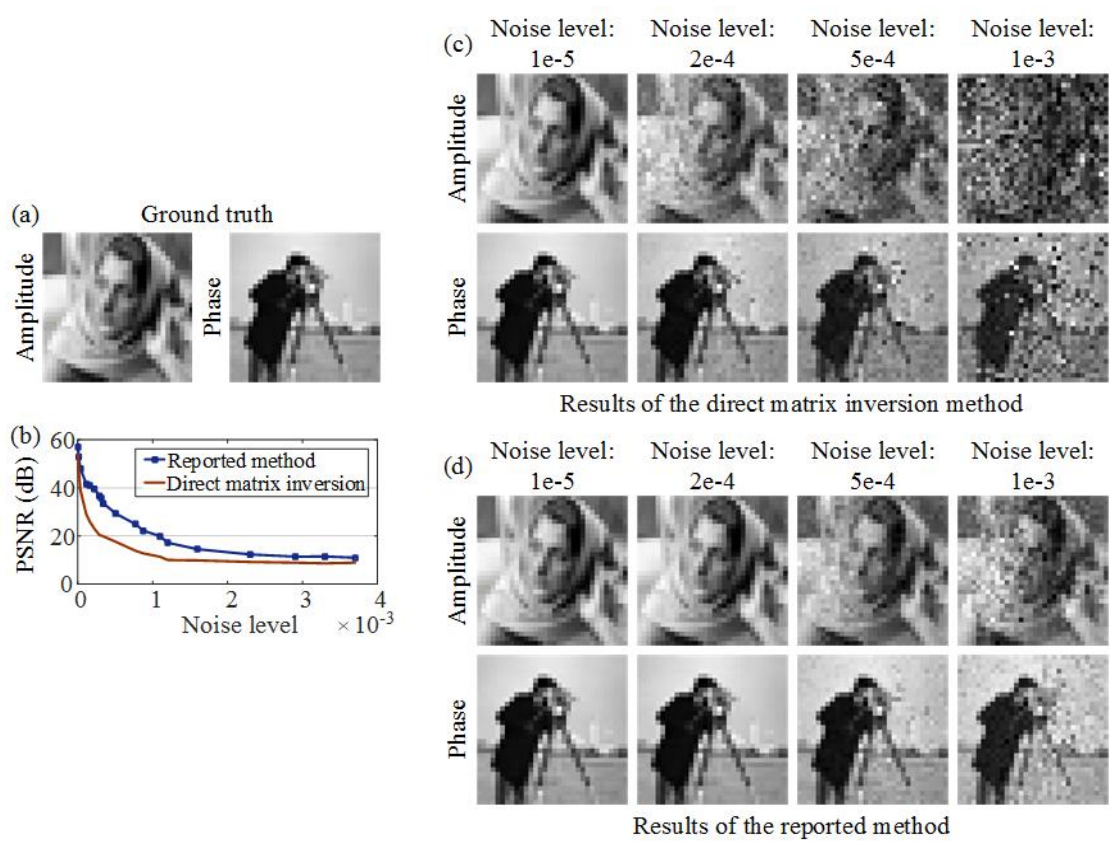

Fig. S2: The reconstructed results under different levels of measurement noise. (a) The ground-truth amplitude and phase maps of a simulated object, with the pixel resolution being $32 \times 32$. (b) The quantitative comparison of reconstruction quality between the reported single-pixel phase retrieval method and the direct matrix inversion method. (c) The reconstructed amplitude and phase maps of the direct matrix inversion method. (d)

The reconstructed amplitude and phase maps of the reported single-pixel phase retrieval method.

During photon detection, measurement noise arises from multiple sources, including dark current and thermal fluctuation. To study the influence of measurement noise on reconstruction quality, Gaussian white noise was added to the simulated measurements, with the noise level quantified by the standard deviation of the Gaussian model. The sampling ratio was fixed at 4. Both the quantitative and qualitative reconstructed results are shown in Fig. S2. We also present the results using the direct matrix inversion method in Ref. [35] as a comparison. We can see that as the noise level increases, the PSNR of the direct matrix inversion method decreases much faster than that of the reported method. As shown in Figs. S2(c) and S2(d), the reported method produces images with less aberrations and more structure details at all noise levels.

## S3: Decrease of measurements' dynamic range

Different from most spatial-domain signals which have a low dynamic range, far-field diffraction patterns rapidly decay radially from the center to the edge by as much as 7 orders of magnitude. As a result, an ultra-high-dynamic-range detector array is required in conventional CDI to record the far-field data. In comparison, the reported technique employs a single-pixel detector to acquire the central DC-only component of the diffraction pattern. Because the DC component does not vary much under different illumination patterns, the measurements' dynamic range is significantly lowered.

We performed a numerical simulation to quantify the decrease of the measurements' dynamic range. The ground-truth amplitude and phase were randomly chosen from the USC-SIPI image database, as shown in Fig. S3(a). We simulated 10 objects with different combinations of these images, and synthesized the corresponding CDI and single-pixel CDI measurements. Figure S3(b) presents the exemplar measurements with the pixel number being $128 \times 128$. The measurement values are displayed on a logarithmic scale. The dynamic range is defined as the ratio between the maximum and minimum values of the measurements. We averaged the dynamic range of all the 10 objects for statistical evaluation, and the results are shown in Fig. S3(c). As pixel number increases, the measurements' dynamic range in conventional CDI increases by orders of magnitude (from $10^7$ to $10^{11}$). This is because the diffraction pattern contains more high-frequency signals with lower intensities. On the contrary, the dynamic range of the reported single-pixel CDI retains the same order of magnitude ($10^0$), with even a small decrease as pixel number increases. The reason may be due to the central DC component representing the overall energy of the light field, which changes less with more pixels. In summary, the above simulation validates that the single-pixel CDI reduces measurements' dynamic range from more than $\sim 10^7$ for conventional CDI to $\sim 10^0$.

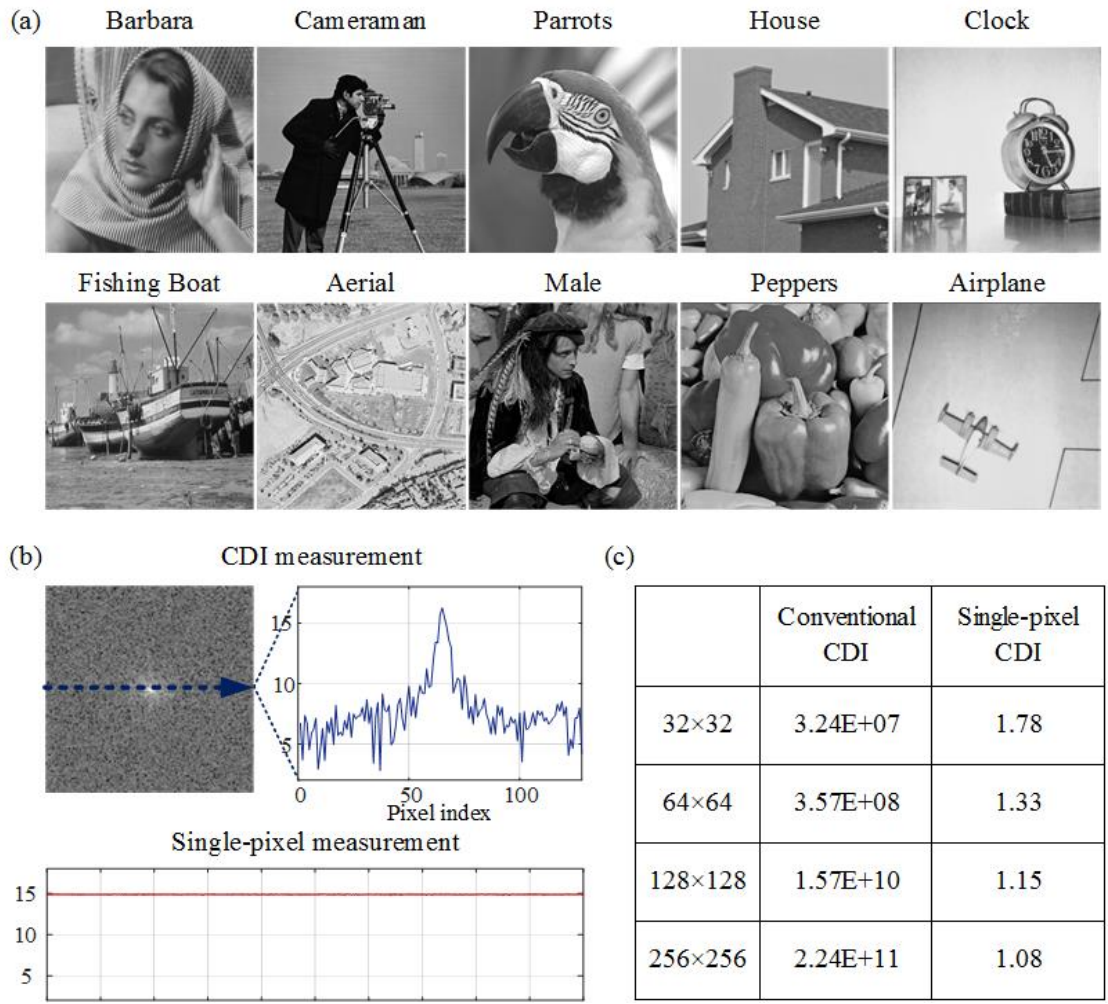

Fig. S3: Comparison of measurements' dynamic range between the conventional CDI

and the single-pixel CDI. (a) The ground-truth amplitude and phase images. (b) The exemplar CDI and single-pixel CDI measurements displayed on a logarithmic scale. (c) The statistical dynamic range of the two techniques under different pixel numbers.

## S4: The influence of detector misalignment on reconstruction

The reported technique employs a single-pixel detector to acquire the DC component of the diffraction pattern. However, the detector may be placed out of the center position in practice, and therefore the measurements correspond to the intensity of other frequency components. We conducted the following simulations to investigate the influence of detector misalignment on the phase-retrieval reconstruction.

In the first simulation, the single-pixel measurement was assumed to be the light intensity of a single frequency around the DC component. Figure S4(a) shows the ground-truth object image and the far-field diffraction pattern. The central 9 entries of the diffraction pattern are labeled using different colors, with the DC component marked in red. The reconstructed results corresponding to the 9 detector positions are presented in Fig. S4(b). The amplitude is correctly reconstructed in every case, but the retrieved phase contains an additional background component.

To eliminate the negative influence of detector misalignment on phase retrieval, we added an additional phase correction process after the reconstruction, by subtracting the background phase from the reconstructed phase. To calibrate the background phase in each experiment setting, we set the object with uniform amplitude and no phase (as shown in Fig. S4(c)), and regarded the recovered phase in this case as the corresponding background phase (as shown in Fig. S4(d)). By subtracting the corresponding background phase in Fig. S4(d) from the reconstructed phase in Fig. S4(b), we obtained the corrected phase maps shown in Fig. S4(e), all of which coincide with the ground-truth phase profile. We note that in practical applications, the calibration of background phase is only implemented once by placing a glass slide as the target and recovering its phase.

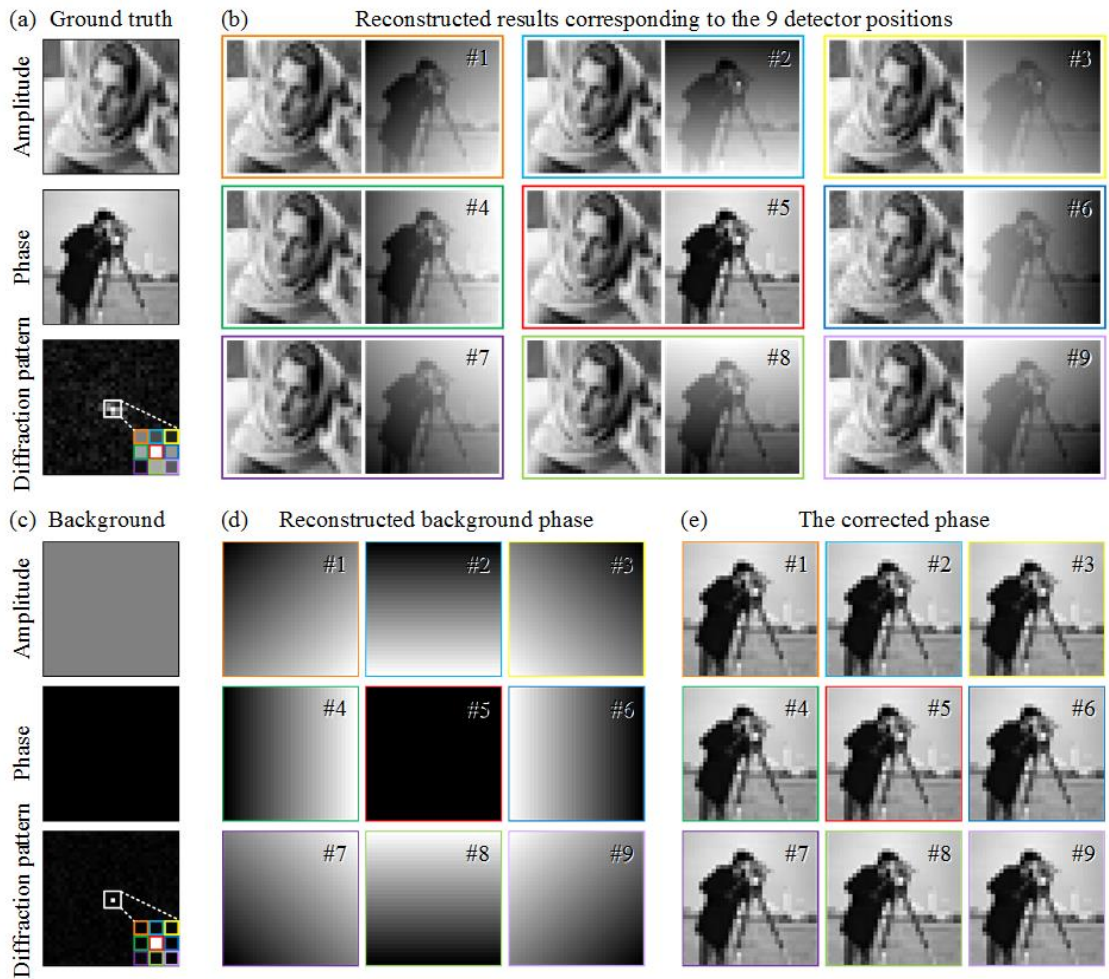

Fig. S4: The reconstructed results corresponding to 9 detector positions around the DC component. (a) The ground-truth object ($32 \times 32$ pixels) and the far-field diffraction pattern. The central 9 entries (from top left to bottom right: #1-#9) are enlarged and labeled with different colors. (b) The uncorrected results corresponding to the 9 detector positions. (c) The simulated object with uniform amplitude and no phase. (d) The reconstructed background phase corresponding to the 9 detector positions. (e) The corrected phase by subtracting the background phase.

In the second simulation, we considered that the single-pixel measurement is the summation of light intensities at multiple spatial frequencies. Four cases were simulated, including detecting the total intensity of one, two, four and nine adjacent frequency components around the DC component. The corresponding reconstructed results are presented in Fig. S5(b). Neither the amplitude nor the phase maps are reconstructed correctly in this case. To avoid the degradation caused by multi-component detection, a pinhole can be placed in front of the single-pixel detector to remove frequencies other than the DC component.

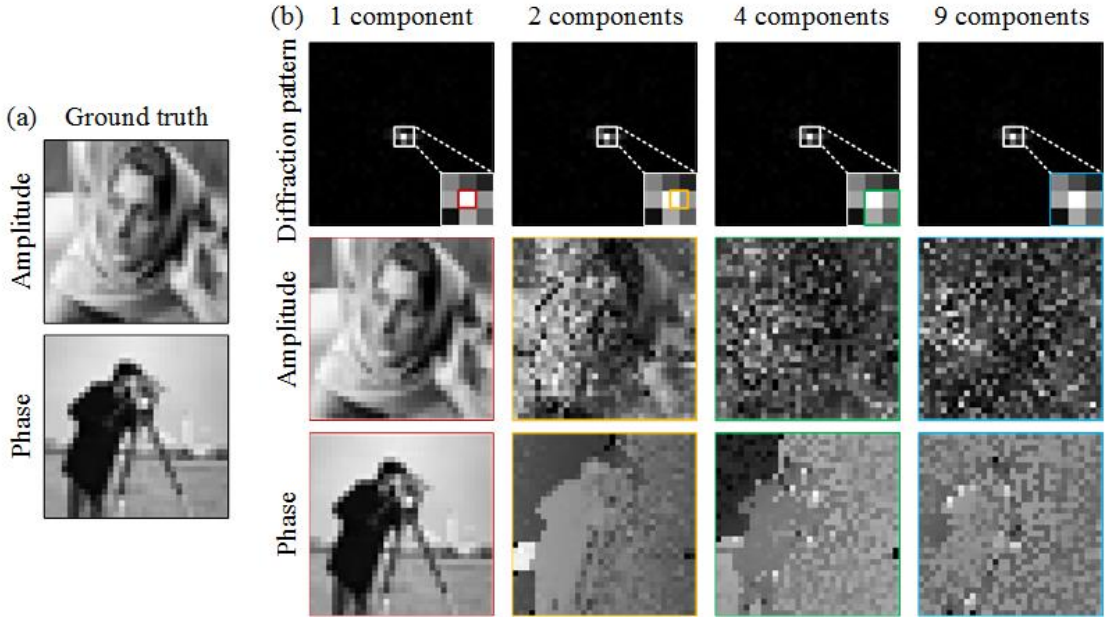

Fig. S5: The reconstructed results in the cases of detecting multiple frequency components. (a) The ground-truth object ($32 \times 32$). (b) The reconstructed results corresponding to detecting the total intensity of one, two, four and nine adjacent frequency components around the DC component.

## S5: The influence of imaging distance on reconstruction

A typical CDI experiment is conducted in the far-field regime, where the imaging distance between the object and detector is much greater than the object size. In such setting, the diffraction pattern approximates to the Fourier transform of the object following Fraunhofer diffraction theory. However, the diffraction pattern can instead follow the Fresnel transform in practice due to a limited imaging distance.

Here we investigate the reconstruction quality of the reported method when the far-field propagation (Fraunhofer diffraction) approximation is not strictly valid. The simulations were set up as follows. The wavelength of illumination was 488 nm. The object size was 2.63mm × 2.63mm, with its ground-truth amplitude and phase images shown in Fig. S6(a). We applied the Fresnel transform to simulate the diffraction pattern. The DC component of the diffraction pattern was regarded as the single-pixel measurement. As shown in Fig. S6(b), the single-pixel detector was placed at different distances from the object along the propagation axis (i.e., the z-direction). We utilized the Fresnel number to characterize the diffraction effect, which is defined as $F = D^2/\lambda z$. Here $D$ is the object size ($D$=2.63mm), $\lambda$ is the wavelength, and $z$ is the distance between the object and the detector. The far-field (Fraunhofer) condition corresponds to a Fresnel number $F<<1$.

The reconstructed results of different imaging distances are presented in Fig. S6(c). The amplitude maps were correctly reconstructed, even though the diffraction does not meet the far-field Fraunhofer condition, but the reconstructed phase maps contain an additional background phase when the imaging distance is small. Similar to the calibration process in dealing with detector misalignment, we calibrated the background phase by setting the object with uniform amplitude and no phase (as shown in Fig. S4(c)), and regarded the recovered phase in each experiment setting as the corresponding background phase (as shown in the fourth row in Fig. S6(c)). By subtracting the background phase from the reconstructed phase, we obtained the corrected phase maps as shown in the bottom of Fig. S6(c). The results show that the reported method works effectively at different measurement distances, even when the detector is not truly in the far field.

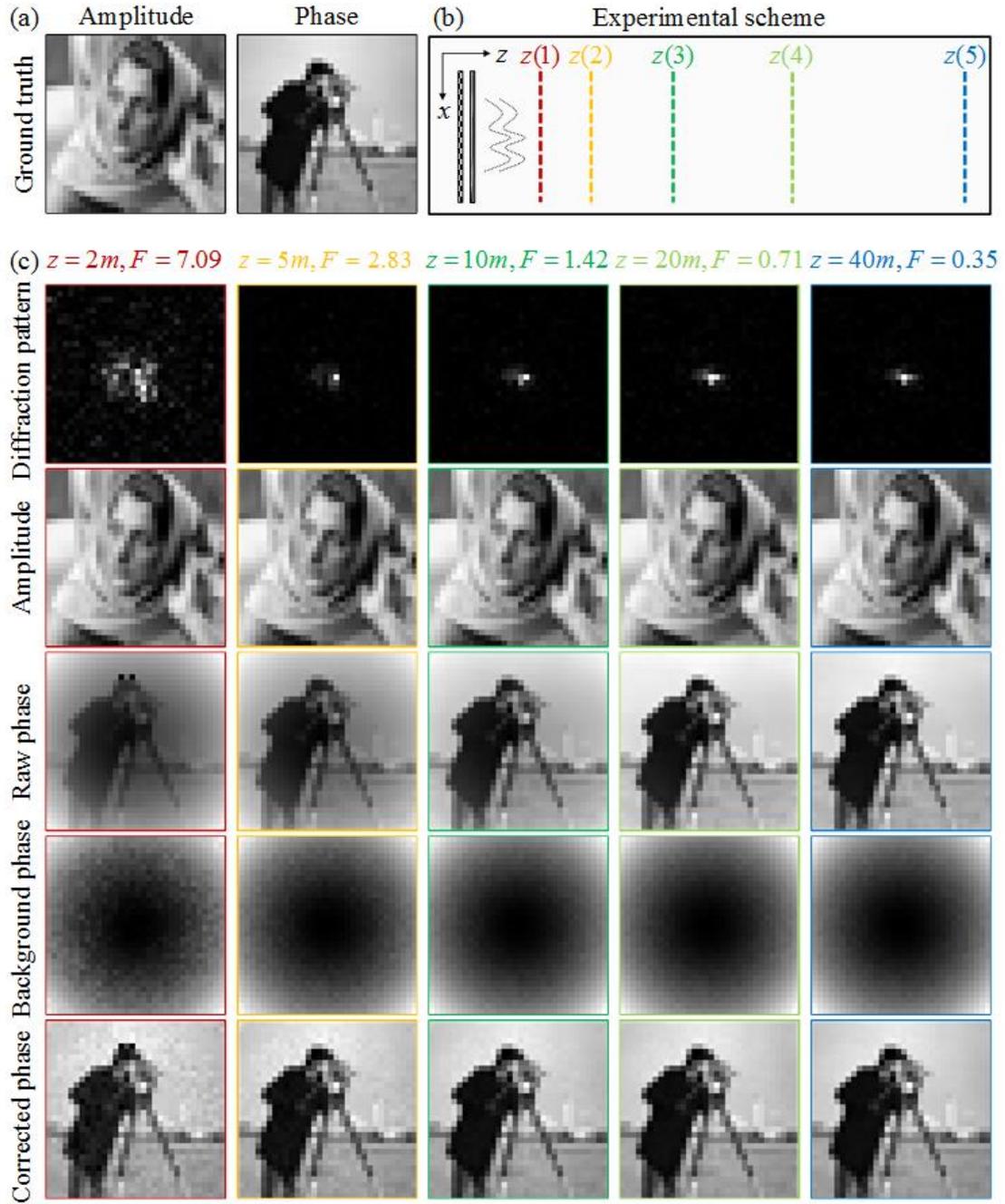

Fig. S6: The reconstructed results at different imaging distances. (a) The ground-truth object (32 × 32). (b) The experimental scheme. The single-pixel detector is placed at different distances from the object along the propagation axis (i.e., the z-direction). (c) The reconstructed results corresponding to different imaging distances.

# S6: Using a SPAD array to reduce sampling ratio and increase imaging efficiency

In our present setting, only the DC component of the diffraction pattern is recorded, and the required sampling ratio is 4. The imaging efficiency can be further improved by acquiring more components of the diffraction pattern. Here we consider replacing the single-pixel detector with a single-photon avalanche diode (SPAD) array. The SPAD array consists of multiple photodetection channels with single-photon sensitivity over a wide working spectrum, enabling multi-component recording of the diffraction pattern. The measurement formation model is described as $I_{k,\gamma} = \left|\sum_{(i,j)} \mathcal{F}[P_k(i,j) \odot O(i,j)] \odot \delta(i - i_\gamma, j - j_\gamma)\right|^2$, where $(i_\gamma, j_\gamma)$ indicates the component frequencies that the SPAD array records in the Fourier plane. Correspondingly, the Fourier-domain update in the reconstruction (Eq. (3) in the Methods section) is revised as $\mathcal{F}[\Psi_{k,\gamma}^{updated}(i,j)] = \mathcal{F}[\Psi_k(i,j)] \cdot [1 - \delta(i - i_\gamma, j - j_\gamma)] + \sqrt{I_k} \frac{\phi_k}{|\phi_k|} \cdot \delta(i - i_\gamma, j - j_\gamma)$.

We implemented simulations to study the imaging efficiency when using a SPAD array, i.e. the relationship between sampling ratio requirement and the channel number of the SPAD array. The parameter settings were the same as those used in the sampling ratio simulations above. The results are summarized in Table S1. We can see that the required sampling ratio is inversely proportional to the number of channels, validating that using multiple photondetection channels is an effective approach to increase imaging efficiency.

Table S1. The required sampling ratio under different channel numbers using a SPAD array

| Channel number | Required sampling ratio |
| --- | --- |
| 4 | 1 |
| 8 | 1/2 |
| 16 | 1/4 |

| 32 | 1/8 |
|---|---|